\documentclass[%
reprint,
amsmath,amssymb,
prb,
floatfix,
]{revtex4-2}

\usepackage[T2A]{fontenc}
\usepackage[cp1251]{inputenc}

\usepackage{color}
\usepackage{graphicx}
\usepackage{dcolumn}
\usepackage{bm}
\usepackage{epstopdf}
\usepackage{float}

\graphicspath{{fig/}}

\begin{document}
	
	\author{Mikhail Malakhov }
	 \affiliation{Institute of Physics, Kazan Federal University, Kazan 420008, Russian Federation}
	 \affiliation{Institut f\"{u}r Theoretische Physik III Ruhr-Universit\"{a}t Bochum, D-44801 Bochum, Germany}
	
	\author{Maxim Avdeev}
	 \affiliation{Institute of Physics, Kazan Federal University, Kazan 420008, Russian Federation}

	\date{\today}
	
	\title{Non-equilibrium  $d-$ wave pair density wave order parameter in superconducting cuprates }

\begin{abstract}
We investigate the nonequilibrium pair density wave order parameter in a simple microscopic model with the ground state $d-$ wave spatially uniform superconductivity. After pushing the system out of equilibrium by a short-time induced order parameter, the system can exhibit robust free nondecaying oscillations of nonuniform superconducting order. In the weak nonequilibrium regime, the frequency of these oscillations is about $2\Delta/\hbar$ and the amplitude can be explained qualitatively by features of equilibrium free energy. In case when the system was taken far from equilibrium, it transits to a non-linear regime where even metastable coexistence of $d-$ wave superconducting and pair density wave gaps are possible.
\end{abstract}

\maketitle


Theoretical research on spatially nonuniform superconducting states - pair density waves (PDW) becomes more relevant due to the latest experiments in cuprates~\cite{Hamidian2016343,Ruan_nature2018,STM_PDW_2019,Arxiv2018HaloPDW}. The PDW is a superconducting state in which the Cooper pairs have a non-zero momentum leading to spatial modulation of the superconducting order parameter. Recently PDW coexisting with uniform $d-$ wave superconductivity was observed in cuprates by scanning tunnel microscopy experiments \cite{Hamidian2016343,Ruan_nature2018,STM_PDW_2019}. Note that the PDW arises in the absence of Zeeman interaction due to an external magnetic field which makes it different from theoretically well described Fulde-Ferrell-Larkin-Ovchinnikov (FFLO)  state~\cite{FF, LO}. 

 Since underdoped cuprates can exhibit a transition to a charge density wave (CDW) order \cite{Ghiringhelli2012821} and competition between CDW and superconductivity were observed in many experiments and theoretical calculations~\cite{Chang2012871,Competing_SC_CDW}, it is very natural to consider that CDW order induces secondary PDW. Just fact of coexisting superconducting (SC) and CDW order parameters give rise to the Cooper pairs with non-zero center-of-mass momentum. But recent experimental and theoretical studies in this area calls into question such a consideration, as it was noted in review ~\cite{Arxiv2019RevPDW}. The appearance of  the PDW order can lead to   break many symmetries, for this reason it gives rise to a variety of induced orders associated  with these broken symmetries. Using Ginzburg-Landau-Wilson formalism it can be shown~\cite{Arxiv2019RevPDW} that different types of PDW can cause  CDW, Ising nematic order, magnetization density wave associated with broken time-reversal and translation symmetries. We also note that some authors associate the pseudogap state to the presence of PDW order~\cite{PDW_pseudogap_Lee_2014,PDW_pseudogap_Lee_2019,PDW_pseudogap_Lee_Taiwan_2019, Pepin_Frac_PDW, Pepin_Frac_PDW2}.

Numerical theoretical calculations for $t-t'-J$ model in cuprates show that despite the fact that usual $d- $ wave SC always corresponds to a global minimum of Free energy at zero temperature, PDW metastable state is very close to it~\cite{RaczkowskiPDW2007,Yang_2009,Spalek_2018,Choubey2017}. Thus, the coexistence of uniform superconductivity and PDW under ordinary conditions is not realized. However, such coexistence becomes possible in vortex halo, where the uniform superconducting gap is locally suppressed by the magnetic field~\cite{YuxuanHalo2018}. In tunneling microscopy experiments \cite{Arxiv2018HaloPDW} $d-$ wave PDW and induced secondary CDW were observed in a vortex halo.

Despite the fact that there is no equilibrium PDW solution for cuprates in simple mean-field microscopic model at reasonable interaction strength and in the absence of other density waves \cite{Loder_PDW2010}, recent works in non-equilibrium dynamics of superconductors \cite{Barankov_Rabi2004,Giannetti_rev_2016, dwave_dyn_2015, SchnyderManske_HiggsLegg,Sentef_CDW_2017, Shimano_rev} motivated us to investigate possible PDW modes when system is driven out of equilibrium.
In works \cite{EreminMuller2018,EreminMuller2019} authors showed that superconducting system can exhibit collective modes due to the subdominant ground state with different symmetry. In work \cite{Photoinduced_eta2019} it was shown that even in  Mott insulator state of the Hubbard model the photoexcited PDW state can exist.

As was noted in \cite{Arxiv2019RevPDW},  at present there is no reliable microscopic theory of PDW in cuprates. This field of research attracts great attention of both theorists and experimenters and now is developing fast. In this rapid communication we show the strong oscillation of PDW order parameter in the simple $t-J$ model
and in the absence   of CDW order.

We start from one-band two-dimensional tight-binding Hamiltonian:
\begin{equation}
H=\sum_{\mathbf{k},\sigma}\varepsilon_{\mathbf{k}} a^\dagger_{\mathbf{k},\sigma}a_{\mathbf{k},\sigma}+H_{int},
\end{equation}
with the following dispersion law 
$
\varepsilon_{\mathbf{k}}= 2 t_1(\cos k_xa + \cos k_ya)
+ 4t_2 \cos k_xa  \cos k_ya 
+ 2 t_3(\cos 2k_xa + \cos 2k_ya) - \mu.
$ 
Here $t_1$, $t_2$ and $t_3$ are effective hopping parameters, $\mu$ is chemical potential.  

We introduce short-range superexchange interaction:
\begin{equation}
H_{int}=-\frac{1}{2}\sum_{\substack{\mathbf{k},\mathbf{k}', \mathbf{k}'',\\
	\sigma,\sigma'}}J_{\mathbf{k''}}a^\dagger_{\mathbf{k},\sigma}a_{\mathbf{k'},\sigma}a^\dagger_{\mathbf{k'-k''},\sigma'}a_{\mathbf{k-k''},\sigma'},
\label{Full_int}
\end{equation}
where $J_{\mathbf{k''}}=2J(\cos k''_xa + \cos k''_ya)$. Than we do mean field decoupling for eq. \ref{Full_int} with three possible combinations of momentum vectors. Term $\mathbf{k'-k''=-k}$ corresponds to uniform superconductivity and terms with $\mathbf{k'-k''=-k\pm} 2\mathbf{q}$ to PDW. Here  2$\mathbf{q}$ is a vector of spatial modulation of order parameter. In this paper we suggest a superconductor which is far from CDW state and we in sake of simplicity neglect interaction terms for CDW. The PDW part of Hamiltonian is the following: 
\begin{equation}
H^\mathrm{PDW}=\sum_{\mathbf{k},s=\pm 1} P^{(s)}_{\mathbf{k}+2s\mathbf{q}}a^\dagger_{\mathbf{k},\uparrow}a^\dagger_{\mathbf{-k}+2s\mathbf{q},\downarrow} + H.C.
\end{equation}
Also due to symmetry, we can write $P^{(+)}_{\mathbf{k+q}}=P^{(-)}_{\mathbf{k-q}}= P_{\mathbf{k}}$.


 If we neglect higher order harmonics, we can write equation of motions in the following form:
\begin{equation}
i\hbar \frac{\partial}{\partial t} \vec{A}  = \hat{M}\vec{A},
\label{eq_mo}
\end{equation} 
where vector 
$
\vec{A}= \begin{bmatrix}
a_{\mathbf{k+q},\uparrow} ,
a_{\mathbf{k-q},\uparrow} ,
a^\dagger_{\mathbf{-k-q},\downarrow} ,
a^\dagger_{\mathbf{-k+q},\downarrow}
\end{bmatrix}^T 
$
and the matrix $\hat{M}$ has the form:
	\begin{equation}
   \hat{M}=	\begin{bmatrix}
	\varepsilon_{\mathbf{k+q}} &0 &\Delta_{\mathbf{k+q}} & P_{\mathbf{k}}\\
	0&\varepsilon_{\mathbf{k-q}} &P_{\mathbf{k}}  &\Delta_{\mathbf{k-q}} \\
	\Delta^*_{\mathbf{k+q}}& P^*_{\mathbf{k}}&-\varepsilon_{\mathbf{-k-q}} & 0\\
	P^*_{\mathbf{k}}&\Delta^*_{\mathbf{k-q}} &0&-\varepsilon_{\mathbf{-k+q}} \\
	\end{bmatrix}.
	\end{equation}
	The excitation spectrum defined as eigenvalues of matrix $\hat{M}$ is the following:
	\begin{equation}
	\Omega^2_{\mathbf{k}, (\pm)}= \frac{1}{2}(E^2_{\mathbf{k+q}} + E^2_{\mathbf{k-q}}+2 |P_{\mathbf{k}}|^2  ) \pm \frac{1}{2}\sqrt{D_{\Omega}}.
	\label{Omega_PDW_SC}
    \end{equation}
Here $E^2_{\mathbf{k}}=\varepsilon^2_{\mathbf{k}} +|\Delta_{\mathbf{k}}|^2 $ is conventional Bogoliubov quasiparticle dispersion,
\begin{equation}
\begin{gathered}
D_{\Omega}	= (E^2_{\mathbf{k+q}}- E^2_{\mathbf{k-q}})^2 +8\mathrm{Re}(\Delta_{\mathbf{k+q}}\Delta_{\mathbf{k-q}} 	 (P^*_{\mathbf{k}})^2)\\
  +4 |P_{\mathbf{k}}|^2 [(\varepsilon_{\mathbf{k+q}} - \varepsilon_{\mathbf{k-q}})^2+(|\Delta_{\mathbf{k+q}}|+|\Delta_{\mathbf{k-q}}|)^2 ]. 
  \end{gathered}
  \end{equation}
Using Green function approach we can write analytical expression for all existing in the model averages.
For PDW anomalous average has the form:
\begin{equation}\label{fPDW}
\begin{gathered}
f^{\mathrm{PDW}}_{\mathbf{k}}=\langle a_{\mathbf{-k+q},\downarrow} a_{\mathbf{k+q},\uparrow}\rangle\\
=\frac{\left[ \varepsilon_{\mathbf{k+q}}\varepsilon_{\mathbf{k-q}}P_{\mathbf{k}}+|P_{\mathbf{k}}|^2P_{\mathbf{k}}-\Delta_{\mathbf{k+q}}\Delta_{\mathbf{k-q}}P^*_{\mathbf{k}}\right]T_1  -  P_{\mathbf{k}} T_2  }{2(\Omega_{\mathbf{k}, +}^2-\Omega_{\mathbf{k}, -}^2)},
\end{gathered}
\end{equation}
where functions $T_1$ and $T_2$ are combinations of hyperbolic tangents:
\begin{equation}
\begin{gathered}
T_1=\frac{\tanh\left(\frac{\beta\Omega_{\mathbf{k}, +}}{2}\right)}{\Omega_{\mathbf{k}, +}} -  \frac{\tanh\left(\frac{\beta\Omega_{\mathbf{k}, -}}{2}\right)}{\Omega_{\mathbf{k}, -}}, \\
T_2=\Omega_{\mathbf{k}, +}\tanh\left(\frac{\beta\Omega_{\mathbf{k}, +}}{2}\right) - \Omega_{\mathbf{k}, -}\tanh\left(\frac{\beta\Omega_{\mathbf{k}, -}}{2}\right).
\end{gathered}
\end{equation}
The order parameters then determined by the following system of mean-field equations:
\begin{equation}
\begin{gathered}
P_{\mathbf{k}}=-\frac{1}{2}\sum_{\mathbf{k'}}J_{\mathbf{k+k' }}\langle a_{\mathbf{-k'+ q},\downarrow}a_{\mathbf{k'+ q },\uparrow}\rangle,\\
\Delta_{\mathbf{k}}=-\frac{1}{2}\sum_{\mathbf{k'}}J_{\mathbf{k+k'}}f^{SC}_{\mathbf{k'}}, \ n=\sum_{\mathbf{k}, \sigma} n_{\mathbf{k}, \sigma}.
\end{gathered}
\end{equation}
The last equation	renormalize chemical potential in dispersion $\varepsilon_{\mathbf{k}}$ for the band with fixed filling $n$. We can also separate kernel of this integral equation and write:
\begin{equation}
\begin{gathered}
\Delta_{\mathbf{k}}=\Delta_{x}\cos k_x + \Delta_{y}\cos k_y - \delta_x\sin k_x - \delta_y\sin k_y,\\
P_{\mathbf{k}}=P_{x}\cos k_x + P_{y}\cos k_y - p_x\sin k_x - p_y\sin k_y.
\end{gathered}
\end{equation}
In \cite{Eremin1998} coexistence of superconductivity and PDW was considered in the strongly correlated regime, more suitable for undoped cuprates, but we see that self-consistent mean field equations differ only on a constant factor.
The free energy of the system can be written as:
\begin{equation}\label{FreeSC_PDW}
\begin{gathered}
F=-\frac{1}{2}\sum_{\mathbf{k},l=\pm}\Omega_{\mathbf{k},l}	+ \frac{1}{2}\sum_{\mathbf{k}} (  \varepsilon_{\mathbf{k-q}}+\varepsilon_{\mathbf{k+q}})\\
+\frac{|\Delta_{d}|^2}{2J} + \frac{|\Delta_{s}|^2}{2J}
+\frac{|P_{d}|^2}{J}+ \frac{|P_{s}|^2}{J},
\end{gathered}
\end{equation}
where we introduced $d-$ and $s-$ components of PDW and SC order parameters as $P_d = P_x - P_y$, $P_s=  P_x + P_y$ and $\Delta_d = \Delta_x - \Delta_y$, $\Delta_s=  \Delta_x + \Delta_y$, respectively. We also assumed, that triplet components $\delta_{x(y)}$ and $p_{x(y)}$ are equal to zero.

A similar model for a PDW and SC was derived in work \cite{Eremin1998}, but in that work, authors concentrated on a CDW and SC coexistence on a $(\pi,\pi)$ vector. Also metastable solutions 
were obtained using a similar technique  in \cite{Loder_PDW2010}. But they did not take into account shift of chemical potential and used unrealistic large interaction strength, for a constant $J>2t_1$. In this case, there are indeed exist PDW solutions, but the value of the superconducting gap is comparable to the bandwidth and critical temperature is much larger than in existing real systems. We use in our calculations interaction constant $J=1.7t_1$, which leads to a reasonable SC gap value. And also we couldn't neglect to shift the chemical potential, which in our case is the same order as the PDW gap. The result of our self-consistent calculation is the full absence of any PDW solutions on different vectors $2\mathbf{q}=(2q_x, 0)$.

Further, in order to get evolution of the system after taking it out of equilibrium, we add time dependence as $ \hat{M} = \hat{M}(t)$ into eq.~\ref{eq_mo}. Such consideration is equivalent to generally accepted methods, introduced in \cite{Barankov_Rabi2004}. But in this model we cannot use pseudospin formalism, and we need to write a complete set of the differential equations for all averages.
Thus the dynamics of existing in system averages can be written as:
\begin{equation}
\label{noneq_dynamic}
\frac{\partial}{\partial t}	K_{mn} =- i \sum_{j} (M_{mj}K_{jn} - M_{jn}K_{mj})/\hbar,
\end{equation}
where $K_{mn}=\langle A_mA_n^\dagger\rangle$.
In particular, PDW average dynamic can be expanded as:
\begin{equation}\label{PDW_diff_eq}
\begin{gathered}
i\hbar \frac{\partial}{\partial t} f_\mathbf{k}^{\mathrm{PDW}} = ( \varepsilon_{\mathbf{k+q}} + \varepsilon_{\mathbf{k-q}})f_\mathbf{k}^\mathrm{PDW} \\
 +P_\mathbf{k}(1- n_{\mathbf{k+q}, \uparrow} - n_{\mathbf{-k+q}, \downarrow}) \\
 +\Delta^*_{\mathbf{k-q}}f^\mathrm{CDW}_\mathbf{k} - \Delta_{\mathbf{k+q}}(f_\mathbf{k}^{\mathrm{CDW}})^{*}.
\end{gathered}
\end{equation}
This equations are similar to the differential equations from~\cite{Sentef_CDW_2017}.
One can see, that PDW can't be excited by applying light pulse, like in pump-probe experiments in cuprates \cite{Shimano_Higgs_dwave_2018, Shimano_LASCO_resp2019,Manske_newmode}. Since there are no equilibrium solutions for any other density wave correlations, we have to induce for a short time PDW order parameter to see the non-equilibrium dynamics of system after it.   Our idea is that proximity of macroscopic cuprate and PDW sample can cause non-equilibrium dynamics of order parameters in cuprates.

\begin{figure}
	\includegraphics[width=0.5\textwidth]{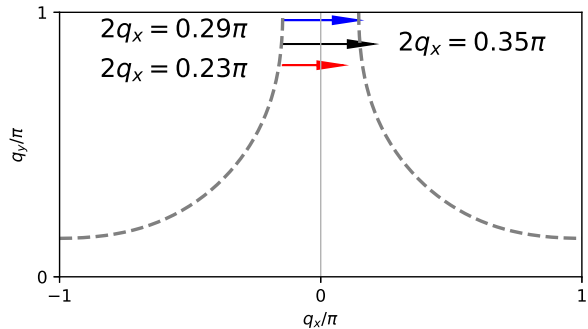}
	\caption{\label{FSq}Fermi surface and PDW modulation vectors $\mathbf{q}=(q_x,0)$ which we used in our calculations. Band parameters: $t_1=100$ meV, $t_2=-0.35t_1$, $t_3=0.05t_1$ with concentration of carriers $n=0.8$.}
\end{figure}

\begin{figure*}
	\includegraphics[width=0.5\textwidth]{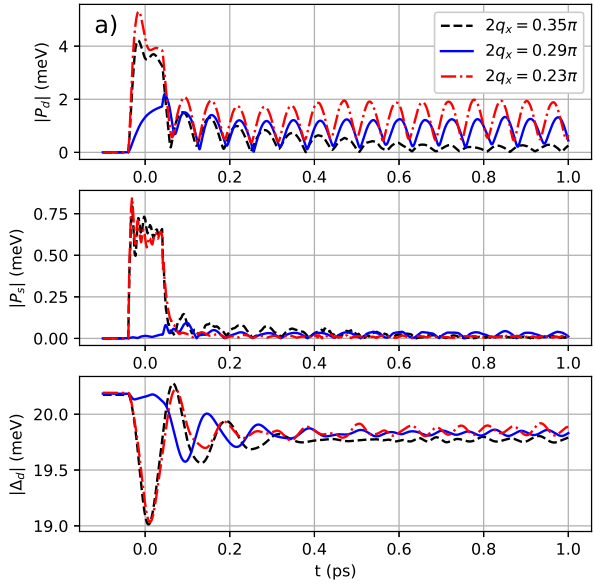}\hfill
	\includegraphics[width=0.5\textwidth]{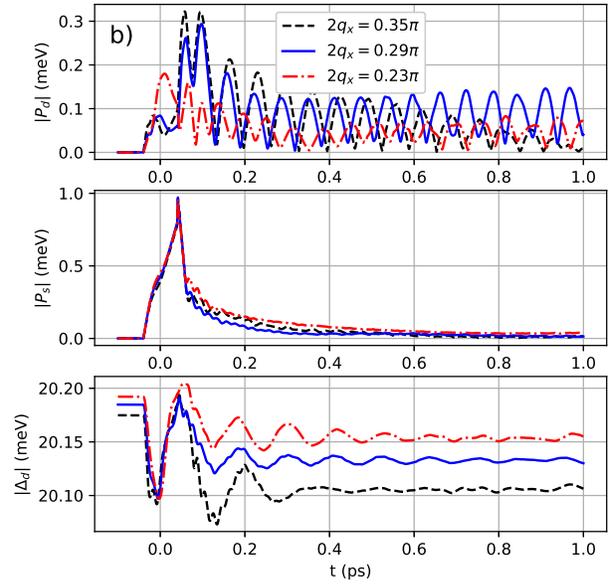}
	\caption{\label{fig_PDW_dyn_w}Non-equilibrium dynamics of PDW and SC order parameters after induced (a) d- wave PDW and (b) s- wave PDW correlations on differrent wave vectors. 
		All band parameters are the same as on Fig.~\ref{FSq}, grid in Brillouin zone is $449\times 449$.  }
\end{figure*}

\begin{figure}[!ht]
	\includegraphics[width=0.5\textwidth]{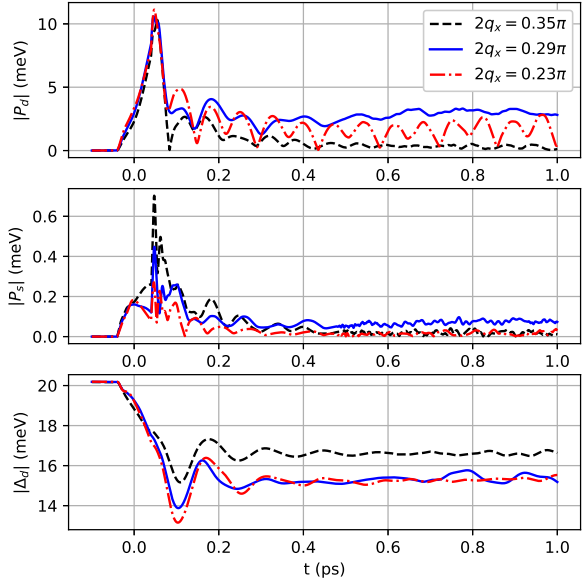}
	\caption{Non-equilibrium dynamics of PDW and SC order parameters after induced d- wave PDW correlations on differrent wave vectors. 	All band parameters are the same as on Fig.~\ref{FSq}.  }
	\label{fig_PDW_dyn_strong}
\end{figure}

In our calculations we induced  external PDW for three different  $2\mathbf{q}$ vector as shown on Fig.~\ref{FSq}. 
We  numerically simulated the system dynamics using differential eq.~(\ref{noneq_dynamic}) with initial conditions corresponding to the equilibrium state. Then, on the time interval  $t=(-0.05,0.05)$ ps we artificially set $P_{\mathbf{k}}$ amplitudes and corresponding averages  $f_\mathbf{k}^\mathrm{PDW}$ defined by equation~(\ref{fPDW}).
 We found that system has two regime of non-equilibrium time evolution  of order parameters. When  the initial amplitude of the induced  order parameter is not very large, we have linear oscillations regime, as it can be seen in Fig.~\ref{fig_PDW_dyn_w}, where we induced PDW with $d-$ and $s-$ wave symmetries. We obtained, that $s-$ wave component is always decaying very fast, while $d-$ wave component exhibit oscillations with slightly different amplitudes depending on value of the vector $\mathbf{q}$.
\begin{figure}
	\includegraphics[width=0.45\textwidth]{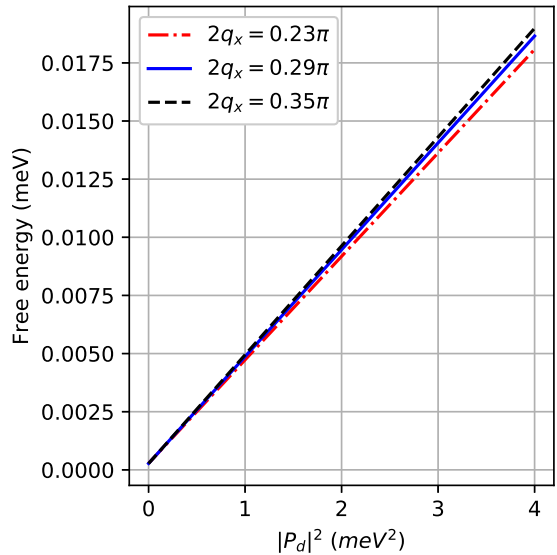}
	\caption{\label{fig:free_energy} The free energy versus amplitude $|P_d|^2$ at a three different value of vector $q$. The SC gap $\Delta_{d}$ is set to its equilibrium value. Other parameters of the system are the same as in Fig.~\ref{FSq}. The normal state free energy is about 0.25 meV.}
\end{figure}
The difference in oscillation amplitudes can be explained qualitatively by analyzing Free energy~(\ref{FreeSC_PDW}). The Free energy dependence on $d-$ wave PDW gap is quadratic ($F \sim |P_d|^2$ see Fig.~\ref{fig:free_energy}) and the oscillations amplitude correlates with its value. Also the free energy for uniform $d-$ wave SC coexisted with $s-$ wave PDW is larger than with $d-$ wave PDW which explains why $s-$ wave oscillations are suppressed.
With a further increase amplitude of the initial induced PDW order parameter, we can see a qualitative change it time behavior (see solid blue line in Fig.~\ref{fig_PDW_dyn_strong}). 
 In this case we also see undamped oscillations of PDW order parameters at wave vector $2q=0.23\pi$, but which are already non-linear. And for $2q=0.29\pi$ system can undergo into a non-equilibrium meta-stable state, where PDW and SC order parameter coexist and show weak time dependence of amplitude at $t>0.4$ ps. 
 This feature is due to the fact that in this mode the nonlinear processes begin to play a decisive role.


One important and open question is how this mode can be excited in experiments, because we need a proximity with already existing PDW order parameter.  The PDW  state can be the ground state in some systems as one dimensional Kondo-Heisenberg model \cite{Berg2010} and $t-J$ model with ring exchange on a triangular lattice \cite{Xu2018} which can be related to cuprates. Also there are recent work  \cite{Venderley_PDW_spin_vall} where PDW state was predicted in hole-doped group VI transition metal dichalcogenides, with spin-valley locked band structure and moderate correlations.  In order to induce the gap, we can also create FFLO state in small system. As it was noted in work \cite{KimFFLO2019} the size of the system is very important in formation of spatially non-uniform superconductivity and the system is stable if the modulation fits the finite system size. So it can occur that PDW or FFLO state is more preferable in small-sized superconductors on a STM tip, than in macroscopic samples. 
For instance, in microscopic calculations, metastable PDW were obtained in small-sized superconductors \cite{RaczkowskiPDW2007,Choubey2017}. Existing microscopic models operate with simulations in on-site representation of small-sized superconductors, usually not larger than $60\times 60$ sites because the complexity of such calculations grow very fast when the system increase. In  work~\cite{Hamidian2016343} the PDW order on top of equilibrium superconductivity was seen in STM experiments, where nano-sized superconducting cuprate flake on a STM tip was placed to measure Josephson current. In addition in work \cite{BabaevPDW} was stressed the importance of surface effects in FFLO state superconductors, which is mostly important in small-sized samples.

The important result of our calculation one can induce oscillations of the PDW order parameter in weak non-equilibrium regime  even in the absence of the CDW order parameter. Wherein time evolution of the PDW exhibit a robust non-decaying oscillation after induced order parameters disappear.  In our calculations we could get PDW gap amplitude oscillations up to 10\% of uniform equilibrium gap. This non-equilibrium PDW gap is always $d-$ wave with a negligibly small admixture of $s-$ wave component, not depending on which gap symmetry was initially induced. Also, such strong oscillations lead only to a small 2\% drop of uniform SC amplitude. The frequency of this oscillation is about $2\Delta^{\mathrm{max}}/\hbar$ and thus coincides with frequency of Higgs mode~\cite{dwave_dyn_2015,Lance_PDW_Higgs_2017}.
 The difference which can be used to distinguish from uniform SC amplitude mode is that PDW oscillations do not decay quickly as it happens in $d-$ wave cuprates \cite{dwave_dyn_2015}.
As it was seen in our calculations, even if only $s-$ wave external PDW was induced, the oscillations in $d-$ wave channel can be excited, when $s-$ wave PDW decay quickly when external gap removed. 
Thus, based on our results, we believe  that manifestation  of the interplay  between PDW and SC order parameters can be detected in systems  far from CDW instability.


We acknowledge stimulating discussions with M. A. M\"{u}ller, A. Kadigrobov and I. Eremin. This work was supported from the project of the state assignment of KFU in the sphere of scientific activities, Grant No. 3.2166.2017/4.6.

\bibliographystyle{apsrev4-2}

\bibliography{MMalakhov_MAvdeev}

\end{document}